\documentclass{article}
\usepackage[a4paper,
top=1.5cm,
bottom=1.5cm,
left=2.5cm,
right=2.5cm,
marginparwidth=1.75cm
]{geometry}

\usepackage{amsmath, amsthm, amssymb}
\usepackage{algorithm, algpseudocodex}
\usepackage{physics}

\usepackage{authblk}
\usepackage[justification=centering, size=small]{caption}
\usepackage{subcaption}

\newtheorem{theorem}{Theorem}

\title{Rotational Encoding of Objective Functions for Quantum Optimization}

\author[1]{Sonia Yang\footnote{sonia.liu.yang@gmail.com}}
\author[2,1]{Ali Al-Bayaty\footnote{ali.albayaty@alnaji-uni.edu.iq; albayaty@pdx.edu}}
\author[3]{Shanyan Chen\footnote{sychen@cnu.edu.cn}}
\author[1]{Marek Perkowski\footnote{h8mp@pdx.edu}}
\affil[1]{\emph{Department of Electrical and Computer Engineering, Portland State University, USA}}
\affil[2]{\emph{Department of AI Engineering, College of Engineering, Al-Naji University, Iraq}}
\affil[3]{\emph{College of Information Engineering, Capital Normal University,
China}}
\date{}

\theoremstyle{definition}

\newtheorem{example}{Example}

\begin{document}
    \maketitle
    \begin{abstract}
         Quantum search algorithms provide computational advantages for a broad class of constraint satisfaction problems, which have been achieved through the construction of problem-specific Boolean-based oracles. Currently, conventional approaches using such oracles in quantum search algorithms for the optimization of objective functions utilize multi-qubit registers that store the objective value and compare it against a repeatedly updated threshold. The arithmetic and comparator circuits that perform these computation and comparison operations are constructed from many quantum reversible gates that extensively contribute to the qubit count and circuit cost. In this paper, we introduce the Angular Rotational Counter (ARC), a novel approach to represent objective functions that can be decomposed as a weighted sum of Boolean functions. ARC encodes the objective as an accumulated sum of quantum rotations onto a single-qubit register instead of a multi-qubit register used in much of the previous literature. We mathematically establish and experimentally validate that these rotations sum to the objective value for all possible solutions and derive the relationship between the resulting measurement probabilities and the objective. This design was applied to construct a quantum oracle for the state assignment problem of finite automata and state machines using a weighted Hamming distance cost function, which is a special case of the quadratic assignment problem. We evaluate this oracle design on 26 benchmarks and compare it with corresponding Boolean-based oracles on four cost metrics. Across these benchmarks, the oracle with ARC produces a mean reduction in the qubit count of 20.8\%, gate count of 89.3\%, circuit depth of 92.4\%, and quantum cost of 90.6\%. These results demonstrate substantial reductions in circuit cost for the evaluated benchmarks and establish rotational encoding as a promising approach for representing objective functions in optimization based on quantum search algorithms.
    \end{abstract}



\section{Introduction}
\label{sec:introduction}
    Quantum search algorithms, including Grover's algorithm~\cite{grover1996quantum, grover1997quantum, grover1998quantum, boyer1998grover,figgatt2017grover,al-bayaty2024diffuser}, quantum walk-based search algorithms~\cite{shenvi2003walk, childs2004walk}, and BHT-QAOA~\cite{al-bayaty2024bht-qaoa,al-bayaty2025bht-qaoa}, have been proven to require less queries than classical exhaustive search. These algorithms are applicable to many constraint satisfaction problems (CSPs) through the construction of problem-specific oracles~\cite{campbell2019csp, tsai2020encoding, hou2020knapsack, perkowski2022grover}. The practicality and performance of these algorithms, however, are critically dependent on the realization and cost of these oracles. Furthermore, additional components are necessary for these oracles when they are designed for optimization problems.

    In previous research on applying quantum search algorithms for combinatorial optimization problems~\cite{tsai2020encoding, hou2020knapsack, durr1996minimum, gilliam2021optimization, ketineni2023esop}, the resulting algorithms were based on repeated, adaptive executions of an oracle-based search algorithm, such as Grover's search algorithm and quantum random walk. Many of these presented oracles~\cite{tsai2020encoding, hou2020knapsack, perkowski2022grover, ketineni2023esop} were entirely Boolean-based and were created based on CSPs. Such oracles use quantum reversible gates to create logical, arithmetic, or comparison operations to determine whether a candidate solution satisfies each constraint. For this reason, these oracles can identify valid solutions, but cannot compare the valid solutions against each other. Therefore, optimization problems were framed as iterated CSPs with a threshold (for the objective function) as an additional constraint. An oracle designed for an iterated CSP computes the objective value using arithmetic operators implemented through reversible logic and encodes the resulting value onto a multi-qubit register. The oracle then employs a comparator to evaluate the computed objective value against a predefined threshold. Depending on the optimization objective, a less-than ($<$) comparator is used for minimization problems, whereas a greater-than ($>$) comparator is used for maximization problems. Hybrid classical-quantum algorithms are constructed by iteratively executing a quantum search algorithm while repeatedly updating the threshold. Although these oracles explicitly compute objective values, resulting in direct access to these values, their implementation requires additional ancilla qubits (temporary working registers) and multiple arithmetic and comparator gates, all of which further increase the circuit cost and depth. These costs depend on the objective function values, their size, and the implementation of the arithmetic and comparator gates. Such costs become more significant as the objective function value increases in size and complexity. An alternative representation that incorporates the information on the objective directly into quantum operations could therefore reduce the cost of such oracles.

    In this work, we introduce the Angular Rotational Counter (ARC), a circuit design that rotationally encodes objective functions that can be decomposed into a weighted sum of Boolean functions. It encodes such objective functions through an accumulated sum of quantum rotational operations. Instead of explicitly computing the objective onto a conventional multi-qubit register as a binary value, ARC associates each term in the objective function with a rotational operation, and the measurement probability of the ARC output depends on the objective value of the candidate solution. Oracles with ARC are executed within a quantum search algorithm, such as Grover's algorithm, to obtain optimized solutions based on the encoded objective function.

    To study the practical applications of our introduced rotational encoding, we investigate quantum oracles with ARC based on a heuristic weighted Hamming distance cost function~\cite{devadas1988mustang, benini1995assignment, tsui1998assignment}, a special case of the quadratic assignment cost function~\cite{ahuja2000qap, loiola2007qap, mermoud2026qap, demicheli1985assignment}, in the context of the state assignment problem~\cite{tsai2020encoding, devadas1988mustang, benini1995assignment, tsui1998assignment, demicheli1985assignment, villa1989nova} for classical and quantum finite automata and state machines~\cite{huang2023encoding, huang2026automata, kumar2025qsm, moore2000automata, kondacs1997automata}. State assignment, also known as ``state encoding," is a fundamental optimization problem in structural automata synthesis and digital design. The assignment of binary encodings to automaton states directly influences the implementation cost and the resulting circuit complexity. Previous work on creating an oracle for finding the minimal encoding of the states of an automaton was discussed in \cite{tsai2020encoding}, which presented a quantum Boolean-based oracle to encode states and inputs based on dependencies on input variables and a repeatedly updated threshold. However, in \cite{tsai2020encoding}, their method contains many expensive multi-qubit gates and restrictions on applicability: the number of states and inputs had to be powers of two, and the automaton had to be completely specified Boolean functions. Our oracle with ARC does not have these restrictions.
    
    The benchmark results demonstrate substantial reductions in quantum circuit resource requirements for the implementations with ARC compared with their corresponding quantum Boolean-based oracles. Across the 26 finite state machine benchmarks, the arithmetic mean reductions are 20.8\% in qubit count, 89.3\% in gate count, 92.4\% in circuit depth, and 90.6\% in circuit cost after transpilation. These results demonstrate the advantages of the oracles with ARC over their Boolean-based counterparts in terms of cost-effectiveness and resource efficiency. Although the observed improvements indicate substantially more compact quantum circuit implementations, they do not, by themselves, establish corresponding reductions in physical execution time or demonstrate an asymptotic quantum speedup. Nevertheless, the substantial reductions in quantum circuit resources motivate further investigation into rotational encodings of objective function values as an alternative to explicit binary encoding. Moreover, the applicability of ARC and its corresponding oracle designs is not limited to the benchmarks and problems considered in this study and may extend to other general classes of problems.
    
    The main contributions of this paper are as follows.
    \begin{itemize}
        \item \textbf{Rotational encoding of decomposable objective functions:} We introduce ARC as a method to represent objective functions that can be decomposed into a weighted sum of Boolean functions. ARC encodes the objective function as an accumulated sum of quantum rotational operations onto a single-qubit register, in contrast to conventional methods that explicitly compute the objective value in binary onto a multi-qubit register. Not only does this reduce the cost of the oracle, but it also introduces a rotational encoding of objective functions that can be utilized for optimization.
        \item \textbf{Mathematical model of ARC:} We demonstrate the accumulated sum of the objective-dependent rotations and formalize the resulting output qubit of ARC for candidate solutions. We also derive the conditions under which the measurement probability is monotonic with respect to the objective function value. This proves that the output qubit of ARC optimizes the measurement probabilities with respect to the objective function value for candidate solutions. 
        \item \textbf{Application and analysis of oracles with ARC for the state assignment problem:} We construct oracles with ARC for the state assignment problem based on a heuristic weighted Hamming distance cost function, which is a special case of the cost function in the quadratic assignment problem. This demonstrates the applicability of ARC to important optimization problems.
        \item \textbf{Evaluation of state assignment oracles:} We compare oracles with ARC against corresponding Boolean-based oracles for state assignment on four circuit cost metrics, namely qubit count, gate count, circuit depth, and Transpilation Quantum Cost (TQC)~\cite{al-bayaty2023gala,al-bayaty2024cala, al-bayaty2026bsa,yang2026decoders}. Our results show cost reductions across all 26 evaluated benchmarks.
    \end{itemize}
    
    The remainder of this paper is organized as follows. Section~\ref{sec:background} reviews the essential background on Grover's algorithm, quantum cost, state assignment, and the weighted Hamming distance cost function. Section~\ref{sec:oracle} presents the proposed Angular Rotational Counter (ARC) circuit with its mathematical representation. Section~\ref{sec:encoding} applies ARC to construct the oracle for the state assignment problem. Section~\ref{sec:experiment} details the experimental methodology used to evaluate such state assignment oracles. Section~\ref{sec:results} presents the results. Section~\ref{sec:discussion} discusses and analyzes the implications of our proposed approach. Lastly, Section~\ref{sec:conclusions} concludes this paper.


\section{Background}
\label{sec:background}

    \subsection{Grover's Algorithm}
    \label{subsec:background-grover}
    
        Classical computing methods to search through an unsorted dataset grow linearly in time complexity with the dataset size. However, in quantum computing, Grover's algorithm~\cite{grover1996quantum, boyer1998grover,figgatt2017grover,al-bayaty2024diffuser} can search through a search space of size $N$ with a query complexity~\cite{ambainis2018querycomplexity} of $O(\sqrt N)$, which is the number of queries the algorithm makes to the oracle. This algorithm searches through an unsorted dataset to find an element that satisfies the oracle, which is the fundamental block for creating quantum and hybrid classical-quantum algorithms for CSP and combinatorial optimization problems. This section presents the general structure of Grover's algorithm. 
    
        Grover's algorithm first applies Hadamard gates to all input qubits, creating a superposition that represents the entire search space. Then, these superposed inputs are run through the oracle. An oracle in Grover's algorithm characterizes the target solutions, which are specific to the problem, by marking them. 
        
\begin{figure}[!htb]
            \centering
            \includegraphics[width=.7\linewidth]{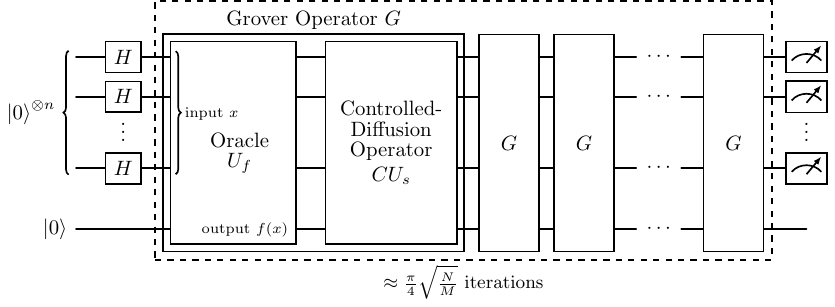}
            \caption{The circuit for Grover's algorithm, which consists of Hadamard gates followed by approximately $\frac{\pi}{4}\sqrt{\frac{N}{M}}$ iterations of Grover's operator $G$, where $N$ is the size of the search space and $M$ is the number of solutions.}
            \label{fig:grover}
        \end{figure}

        A common type of oracle is a Boolean oracle $U_f$~\cite{grover1996quantum, figgatt2017grover, al-bayaty2024diffuser}, which outputs $\ket{1}$ if the inputs are a solution and outputs $\ket{0}$ if not. The output of $U_f$ is conceptually represented by a Boolean function $f(x)$ and is realized using quantum reversible gates~\cite{perkowski2022grover, toffoli1980reversible, peres1985reversible}. After the oracle $U_f$, the controlled-diffusion operator $CU_s$~\cite{al-bayaty2024diffuser} is applied, which increases the probability of measuring inputs that correspond to solutions, based on the Boolean outputs of $U_f$. The combination of the oracle $U_f$ followed by the controlled-diffusion operator $CU_s$ is called the Grover operator $G$. After approximately $\frac{\pi}{4}\sqrt{\frac{N}{M}}$ iterations of $G$ (where $N=2^n$ is the size of the search space and $M$ is the number of solutions), the probability of measuring the target solution approaches 1. The complete construction of Grover's algorithm with a Boolean oracle $U_f$ and controlled-diffusion operator $CU_s$ is illustrated in Fig.~\ref{fig:grover}. There are other types of quantum oracles, such as the phase oracle~\cite{grover1996quantum, figgatt2017grover, al-bayaty2024diffuser, al-bayaty2024bht-qaoa} (which uses the standard diffusion operator $U_s$, not the controlled diffusion operator). In this paper, a modified version of the Boolean oracle is utilized.
    
    \subsection{Quantum Cost Metrics}
    \label{subsec:background-qc}
        The cost of a circuit can be calculated in many ways, such as qubit count, gate count, depth, and Transpilation Quantum Cost (TQC)~\cite{al-bayaty2023gala, al-bayaty2024cala, al-bayaty2026bsa, yang2026decoders}. All of these metrics will be used for evaluation. 

        The qubit count of a circuit is the number of qubits it contains. Because of the limited size and noise of quantum computers, it is important to have circuits with a small number of qubits and gates to prevent decoherence~\cite{shor1995decoherence, krojanski2006decoherence}. The gate count is the total number of gates, and it is usually calculated after all gates are decomposed into native gates, which are the supported operations the quantum hardware runs on. However, it is important to note that some gates are more expensive than other gates. Depth is the critical longest path of the gates along a transpiled circuit. The Transpilation Quantum Cost (TQC) is a metric that calculates the quantum cost of a circuit after transpilation. The TQC of a circuit is determined by Eq.~\eqref{eq:tqc}, where $N_1$ is the number of native single-qubit gates, $N_2$ is the number of native double-qubit gates, $XC$ is the number of SWAP gates~\cite{barenco1995gates, li2019mapping}, and $D$ is the depth.
        \begin{equation}
            \label{eq:tqc}
            \text{TQC} = N_1 + N_2 + XC + D
        \end{equation}

    \subsection{Finite Automata and State Assignment}
    \label{subsec:background-encoding}
    
        The state assignment problem for finite automata and state machines~\cite{tsai2020encoding, devadas1988mustang, benini1995assignment, tsui1998assignment, demicheli1985assignment, villa1989nova} is an important problem that greatly impacts the cost of realizing automata in classical and quantum technologies. There is both the abstract theory of automata~\cite{moore2000automata, kondacs1997automata} and the structural theory of automata~\cite{huang2023encoding, huang2026automata, kumar2025qsm}. This problem is one of the most fundamental problems in the structural theory of automata, and it is stated as follows.
        \begin{itemize}
            \setlength{\itemsep}{0px}
            \item \textbf{Given:} An automaton with states $S_1,S_2,\dots,S_n$, where $n \in \mathbb{N}$ is the number of states.
        \item \textbf{Goal:} Find the state encodings $E_1, E_2,\dots, E_n$ so that the physical realization of the automaton is close to minimal in cost, where $E_i \in \{0,1\}^m$ is the encoding of state $S_i$ and $m\in\mathbb{N}$. 
        \item \textbf{Constraint:} For all $i\neq j$, $E_i \neq E_j$. In other words, the state encodings must be distinct for each state.
        \end{itemize}
    
        The state encodings affect the cost of the resulting circuit. The optimal encoding is affected by the hardware technology.

    \subsection{Weighted Hamming Distance Cost Function}
    \label{subsec:background-cost}

    In the literature~\cite{devadas1988mustang, benini1995assignment, tsui1998assignment}, an effective heuristic cost function for state assignment is one based on a weighted sum of the Hamming distances (HDs) between state encodings. Algorithms minimize this cost function to approximate the solution. Hamming distance is defined in Eq.~\eqref{eq:hd}, where $E_{i,k}$ is the $k$th bit in the encoding $E_i$ and $m$ is the length of the encodings. The weighted Hamming distance cost function is defined in Eq.~\eqref{eq:whd-cost}, where $w_{i,j} \geq 0$ is the weight between the encodings $E_i$ and $E_j$.
    \begin{align}
    \label{eq:hd}
        HD(E_i,E_j) &= \sum_{k=1}^m\left(E_{i,k} \oplus E_{j,k}\right) \\
    \label{eq:whd-cost}
        C &= \sum_{i,j\in\{1,2,\dots,n\},i<j}w_{i,j} HD(E_i,E_j) 
    \end{align}

    There are different methods of assigning weights. One approach is to let $w_{i,j}$ equal the total number of transitions between state $S_i$ and $S_j$. This approach is simpler and suitable for counters, autonomous machines, and flowcharts~\cite{huang2023encoding}. It can be applied to both classical and quantum automata. Another approach is to let $w_{i,j}$ be the sum of the number of literals in the transition function from state $S_i$ to $S_j$ and the transition function from state $S_j$ to $S_i$. A standard approach is to use the minimal Sum of Products (SOP) expression~\cite{quine1952sop, mccluskey1956sop, brayton1982espresso} for a classical automaton realization and an Exclusive-or Sum of Products (ESOP)~\cite{song1993exorcism, song1997exorcism, mishchenko2001exorcism, ketineni2023esop} expression for a quantum automaton realization. These expressions can be calculated using classical or quantum algorithms.


\section{Quantum Oracles with Angular Rotational Counters}
\label{sec:oracle}
    
    The oracle is a fundamental part of many quantum search algorithms, such as Grover's algorithm~\cite{grover1996quantum, boyer1998grover,figgatt2017grover,al-bayaty2024diffuser} and BHT-QAOA~\cite{al-bayaty2024bht-qaoa,al-bayaty2025bht-qaoa}. This section introduces the concept of an Angular Rotational Counter (ARC) and its application in quantum oracles for transforming search algorithms into optimization ones.

    This method is designed for combinatorial optimization problems with objective functions of the form expressed in Eq.~\eqref{eq:objective}, where $m\in\mathbb{N}$, $w_i\geq 0$ is a weight, and $f_i(x)$ is a Boolean function. This covers a broad class of objective functions, and many functions, such as the weighted Hamming distance cost function, can be decomposed into this structure.
    \begin{equation}
    \label{eq:objective}
        C(x) = \sum_{i=1}^mw_if_i(x)
    \end{equation}

    \subsection{Angular Rotational Counter (ARC)}
    \label{subsec:oracle-arc}

        In this paper, we introduce the Angular Rotational Counter (ARC) as an approach for computing objective functions in combinatorial optimization problems using quantum circuits. The ARC circuit computes a rotational encoding of Eq.~\eqref{eq:objective}.

        \begin{figure}[!htb]
            \centering
            \includegraphics[width=0.8\linewidth]{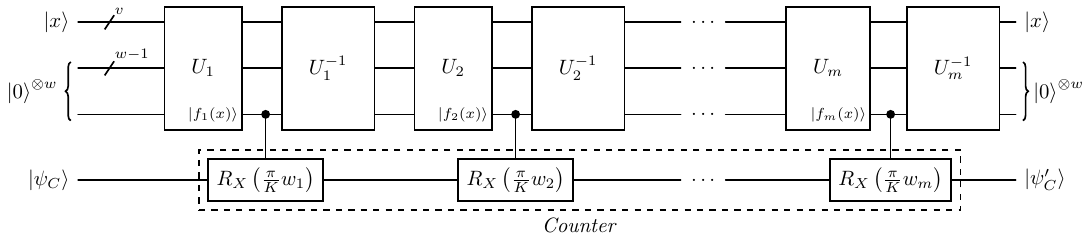}
            \caption{The Angular Rotational Counter (ARC) circuit.}
            \label{fig:arc}
        \end{figure}
        
        The structure of the ARC circuit is illustrated in Fig.~\ref{fig:arc}. The input is the candidate solution $x$, which is a binary vector. For an objective function $C(x)$ described by Eq.~\eqref{eq:objective}, each function $f_i(x)$ is realized in the circuit on a single qubit. This qubit then acts as a control for a rotational gate with angle $\frac{\pi}{K}w_i$ radians, where $K>0$, that is applied to the output qubit with an initial state of  $\ket{\psi_{C}}=\alpha_{C}\ket0+\beta_{C}\ket1$. Note that both controlled-$RX$ and controlled-$RY$ gates can be used in ARC, but the axis of rotation must be consistent throughout the circuit. Controlled-$RX$ gates are used in this paper. Each controlled-$RX\left(\frac{\pi}{K}w_i\right)$ represents an increment of $RX\left(\frac{\pi}{K}w_if_i(x)\right)$, since the rotation of $\frac{\pi}{K}w_i$ is applied if $f_i(x)=1$, and no rotation is applied if $f_i(x)=0$. The accumulated angle of rotation after all controlled-$RX$ gates is
        $$\sum_{i=1}^m \left(\frac{\pi}{K}w_if_i(x)\right) = \frac{\pi}{K}\sum_{i=1}^m w_if_i=\frac{\pi}{K}C(x).$$

    \subsection{Mathematical Correctness of ARC}
        
        This section mathematically shows that the ARC circuit rotationally encodes the objective function $C(x)$ for a superposition of inputs. Theorem~\ref{thm:rotation} proves that the effect of the ARC circuit on the output qubit is equivalent to applying a quantum rotational operation with an angle proportional to $C(x)$ for a single candidate solution $x$ (the input), and it thus demonstrates that the ARC circuit rotationally encodes $C(x)$. Note that the ARC circuit is implemented using controlled-$RX$ gates, but the outcome of the circuit is equivalent to applying a single $RX$ gate to the output qubit with an angle proportional to $C(x)$. Theorem~\ref{thm:superposition} proves that, for a superposed input, the final state of the input qubits and output qubit is an entangled superposition of the corresponding inputs and outputs for each basis state.
        
        \begin{theorem}
        \label{thm:rotation}
            For a single candidate solution $x$ as the input, the final quantum state of the output qubit of an ARC circuit $\ket{\psi_C'(x)}$ is determined by Eq.~\eqref{eq:arc-output}, where the initial state of the output qubit is $\ket{\psi_C}$ and $K>0$.
            \begin{equation}
            \label{eq:arc-output}
                \ket{\psi_C'(x)} 
                = RX\left(\frac{\pi}{K}C(x)\right)\ket{\psi_{C}}
            \end{equation}
        \end{theorem}
        \begin{proof}
            Let $\ket{\psi_{C,i}}=\alpha_{C,i}\ket0 + \beta_{C,i}\ket1$ be the state of the output qubit before the $i$\textsuperscript{th} controlled-$RX$ gate. The result of applying the $i$\textsuperscript{th} controlled-$RX$ gate on the control qubit with state $\ket{f_i(x)}$ and target qubit with state $\ket{\psi_{C,i}}$ is expressed below, where $\overline{f_i(x)}$ represents the Boolean negation of $f_i(x)$.
            \footnotesize
            \begin{equation*}
                CRX\left(\frac{w_i\pi}{K}\right) \left(\ket{f_{i}}\otimes\ket{\psi_{C,i}}\right) =
                \begin{bmatrix}
                1 & 0 & 0 & 0 \\
                0 & 1 & 0 & 0 \\
                0 & 0 & \cos\left(\frac{\pi}{2K}w_i\right) & -i\sin\left(\frac{\pi}{2K}w_i\right) \\
                0 & 0 & -i\sin\left(\frac{\pi}{2K}w_i\right) & \cos\left(\frac{\pi}{2K}w_i\right)
                \end{bmatrix}
                \left(
                \begin{bmatrix}
                    \overline{f_i(x)} \\
                    f_i(x)
                \end{bmatrix}
                \otimes
                \begin{bmatrix}
                    \alpha_{C,i} \\
                    \beta_{C,i} \\
                \end{bmatrix}
                \right) =
                \begin{bmatrix}
                    \alpha_{C,i}\overline{f_i(x)} \\
                    \beta_{C,i}\overline{f_i(x)} \\
                    \gamma_{C,i}f_i(x) \\
                    \delta_{C,i}f_i(x)
                \end{bmatrix}
            \end{equation*}
            \begin{align*}
                \gamma_{C,i}
                &= \alpha_{C,i}\cos\left(\frac{\pi}{2K}w_i\right) - i\beta_{C,i}\sin\left(\frac{\pi}{2K}w_i\right) \\
                \delta_{C,i}
                &= \beta_{C,i}\cos\left(\frac{\pi}{2K}w_i\right) - i\alpha_{C,i}\sin\left(\frac{\pi}{2K}w_i\right)
            \end{align*}
            \begin{align}
                CRX\left(\frac{w_i\pi}{K}\right) \left(\ket{f_{i}}\otimes\ket{\psi_{C,i}}\right)
                &= 
                \alpha_{C,i}\overline{f_i(x)} \ket{00} +
                \beta_{C,i}\overline{f_i(x)} \ket{01} + 
                \gamma_{C,i}f_i(x) \ket{10} +
                \delta_{C,i}f_i(x) \ket{11} \nonumber \\
                &= 
                \alpha_{C,i}\overline{f_i(x)} \left(\ket{0} \otimes \ket{0}\right) +
                \beta_{C,i}\overline{f_i(x)} \left(\ket{0} \otimes \ket{1}\right) + 
                \gamma_{C,i}f_i(x) \left(\ket{1} \otimes \ket{0}\right) +
                \delta_{C,i}f_i(x) \left(\ket{1} \otimes \ket{1}\right) \nonumber \\
                &= 
                \left(\ket{0} \otimes \alpha_{C,i}\overline{f_i(x)} \ket{0}\right) +
                \left(\ket{0} \otimes \beta_{C,i}\overline{f_i(x)} \ket{1}\right) + 
                \left(\ket{1} \otimes \gamma_{C,i}f_i(x) \ket{0}\right) +
                \left(\ket{1} \otimes \delta_{C,i}f_i(x) \ket{1}\right) \nonumber \\
                &= 
                \left(\ket{0} \otimes\left(\alpha_{C,i}\overline{f_i(x)} \ket{0} + \beta_{C,i}\overline{f_i(x)} \ket{1}\right)\right) + 
                \left(\ket{1} \otimes \left(\gamma_{C,i}f_i(x) \ket{0} + \delta_{C,i}f_i(x) \ket{1}\right)\right) \nonumber \\
                &= 
                \overline{f_i(x)}\left(\ket{0} \otimes\left(\alpha_{C,i}\ket{0} + \beta_{C,i}\ket{1}\right)\right) + 
                f_i(x)\left(\ket{1} \otimes \left(\gamma_{C,i}\ket{0} + \delta_{C,i}\ket{1}\right)\right) \nonumber \\
                &=
                \begin{cases}
                    \ket{0} \otimes\left(\alpha_{C,i}\ket{0} + \beta_{C,i}\ket{1}\right) & \text{if }f_i(x)=0 \\
                    \ket{1} \otimes \left(\gamma_{C,i}\ket{0} + \delta_{C,i}\ket{1}\right) & \text{if }f_i(x)=1
                \end{cases} \nonumber \\
            \label{eq:rx-output}
                CRX\left(\frac{w_i\pi}{K}\right) \left(\ket{f_{i}}\otimes\ket{\psi_{C,i}}\right)&=
                \begin{cases}
                    \ket{0} \otimes \ket{\psi_{C,i}} & \text{if }f_i(x)=0 \\
                    \ket{1} \otimes RX\left(\frac{\pi}{K}w_i\right)\ket{\psi_{C,i}} & \text{if }f_i(x)=1
                \end{cases}
            \end{align}
            \normalsize
    
            From Eq.~\eqref{eq:rx-output}, it can be found that the control qubit with state $\ket{f_i(x)}$ remains unchanged after the controlled-$RX$ gate, and Eq.~\eqref{eq:rx-output-qubit} can then be derived (where $RX(0)=I$, the identity operator). Eq.~\eqref{eq:rx-output-qubit} proves that the $i$\textsuperscript{th} controlled-$RX$ gate encodes the $w_if_i$ term.
            \begin{equation}
            \label{eq:rx-output-qubit}
                \psi_{C,i+1} = RX\left(\frac{\pi}{K}w_if_i(x)\right)\ket{\psi_{C,i}}
            \end{equation}        

            Applying Eq.~\eqref{eq:rx-output-qubit} for all values of $i\in\{1,2,\dots,m\}$ results in the following equation. 
            $$\ket{\psi_C'(x)} 
            = \left(\prod_{i=1}^m RX\left(\frac{\pi}{K}w_if_i(x)\right)\right)\ket{\psi_{C}}$$
            
            The composition of rotations is equal to a single complete rotation, i.e., $RX(\theta)\cdot RX(\phi)=RX(\theta+\phi)$. Therefore,  $\ket{\psi_C'(x)} 
            = RX\left(\sum_{i=1}^m\frac{\pi}{K}w_if_i(x)\right)\ket{\psi_{C}}$. Using the definition of $C(x)$ from Eq.~\eqref{eq:objective}, $\ket{\psi_C'(x)}=RX\left(\frac{\pi}{K}C(x)\right)\ket{\psi_{C}}$.
        \end{proof}
        
        \begin{theorem}
        \label{thm:superposition}
            For a superposition of candidate solutions $\ket{\psi_x} = \sum_x c_x\ket{x}$ as the input, the final quantum state of the entangled input and output qubits of an ARC circuit $\ket{\psi'_C}$ is determined by Eq.~\eqref{eq:arc-output-superposition}, where the initial state of the output qubit is $\ket{\psi_C}$.
            \begin{equation}
            \label{eq:arc-output-superposition}
                \ket{\psi'_C} 
                = \sum_x c_x\left(\ket{x} \otimes RX\left(\frac{\pi}{K}C(x)\right)\ket{\psi_{C}}\right)
            \end{equation}
        \end{theorem}
        \begin{proof}
            Theorem~\ref{thm:rotation} states that for a single candidate solution $x$, $\ket{\psi_C'(x)}=RX\left(\frac{\pi}{K}C(x)\right)\ket{\psi_{C}}$, and thus the final state of the input and output qubits is $\ket{x\psi_C'(x)}=\ket{x}\otimes RX\left(\frac{\pi}{K}C(x)\right)\ket{\psi_{C}}$. Therefore, based on linearity, the final state is $\sum_xc_x\ket{x\psi_C'(x)} = \sum_x c_x \left(\ket{x}\otimes RX\left(\frac{\pi}{K}C(x)\right)\ket{\psi_{C}}\right)$ for the superposed input $\ket{\psi_x}$.
        \end{proof}

     \subsection{Relationship Between Objective and Measurement Probabilities}

        The ARC circuit optimizes the measurement probability of $1$ with respect to the objective function $C(x)$. It performs either maximization, as proven in Theorem~\ref{thm:maximization}, or minimization, as proven in Theorem~\ref{thm:minimization}, depending on $\ket{\psi_C}$.

        \begin{theorem}
        \label{thm:maximization}
            If $\ket{\psi_C}=\ket0$, then the measurement probability of $1$ in $\ket{\psi'_C}$ monotonically increases on an interval $(0,K)$, where $K>0$, with respect to $C(x)$.
        \end{theorem}
        \begin{proof}
            From Theorem~\ref{thm:rotation}, the final state $\ket{\psi_C'(x)}$ of the output qubit with $\ket{\psi_C}=\ket0$ is $$\ket{\psi_C'(x)}=RX\left(\frac{\pi}{K}C(x)\right)\ket0=\cos\left(\frac{\pi}{2K}C(x)\right)\ket{0}-i\sin\left(\frac{\pi}{2K}C(x)\right)\ket{1}.$$ 
            The probability amplitude of $\ket{1}$ for $\ket{\psi_C}$ is $\beta_C'=-i\sin\left(\frac{\pi}{2K}C(x)\right)$, which has a period of $\frac{2\pi}{\frac{\pi}{2K}}=4K$. The measurement probability of $1$ is $|\beta_C'|^2=\left(\sin\left(\frac{\pi}{2K}C(x)\right)\right)^2$. Because $\sin\left(\frac{\pi}{2K}C(x)\right)$ is positive and monotonically increases on the interval $(0,K)$ with respect to $C(x)$, $|\beta_C'|^2$ is monotonically increasing on the interval $(0,K)$ with respect to $C(x)$.
        \end{proof}

        \begin{theorem}
        \label{thm:minimization}
            If $\ket{\psi_C}=\ket1$, then the measurement probability of $1$ in $\ket{\psi'_C}$ monotonically decreases on an interval $(0,K)$, where $K>0$, with respect to $C(x)$.
        \end{theorem}
        \begin{proof}
            From Theorem~\ref{thm:rotation}, the final state $\ket{\psi_C'(x)}$ of the output qubit with $\ket{\psi_C}=\ket1$ is $$\ket{\psi_C'(x)}=RX\left(\frac{\pi}{K}C(x)\right)\ket1=-i\sin\left(\frac{\pi}{2K}C(x)\right)\ket{0}+\cos\left(\frac{\pi}{2K}C(x)\right)\ket{1}.$$
            The probability amplitude of $\ket{1}$ for $\ket{\psi_C}$ is $\beta_C'=\cos\left(\frac{\pi}{2K}C(x)\right)$, which has a period of $4K$. The measurement probability of $1$ is $|\beta_C'|^2=\left(\cos\left(\frac{\pi}{2K}C(x)\right)\right)^2$. Because $\cos\left(\frac{\pi}{2K}C(x)\right)$ is positive and monotonically decreases on the interval $(0,K)$ with respect to $C(x)$, $|\beta_C
            '|^2$ is monotonically decreasing on the interval $(0,K)$ with respect to $C(x)$.
        \end{proof}

\begin{figure}[!htb]
            \centering
            \begin{subfigure}{0.48\linewidth}
                \centering
                \includegraphics[width=0.8\linewidth]{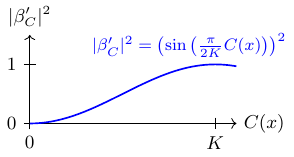}
                \caption{If $\ket{\psi_C}=\ket0$, $|\beta_C'|^2$ increases with respect to the objective $C(x)$ on $(0,K)$. This corresponds to maximization.}
                \label{subfig:increasing-probability}
            \end{subfigure}
            \hfill
            \begin{subfigure}{0.48\linewidth}
                \centering
                \includegraphics[width=0.8\linewidth]{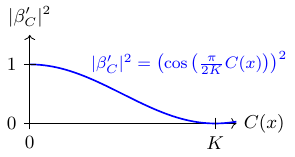}
                \caption{If $\ket{\psi_C}=\ket1$, $|\beta_C'|^2$ decreases with respect to the objective $C(x)$ on $(0,K)$. This corresponds to minimization.}
                \label{subfig:decreasing-probability}
            \end{subfigure}
            \caption{The relationship between $|\beta_C'|^2$ and $C(x)$ on the interval $(0,K)$, where $|\beta_C'|^2$ is the probability of measuring 1 in the state $\ket{\psi'_C}$ and $C(x)$ is the objective function.}
            \label{fig:probability-vs-objective}
        \end{figure}

        The graph of the measurement probability of 1 in $\ket{\psi'_C(x)}$ is illustrated in Fig.~\ref{subfig:increasing-probability} if $\ket{\psi_C}=\ket0$ and Fig.~\ref{subfig:decreasing-probability} if $\ket{\psi_C}=\ket1$. Theorem~\ref{thm:maximization} proves that if $\ket{\psi_C}=\ket0$, as the objective of a solution increases, the measurement probability increases. Therefore, if $K$ is set to the maximum possible objective, then for all solutions, those with a higher objective value are selected with a higher probability after measurement. Therefore, ARC maximizes the measurement probabilities of solutions. In the case where $\ket{\psi_C}=\ket1$, as proven in Theorem~\ref{thm:minimization}, the opposite effect takes place: as the objective of a solution increases, the measurement probability decreases. Hence, ARC minimizes the measurement probabilities of solutions instead.
        
        On the one hand, Boolean-based oracles can distinguish between the inputs that are solutions and those that are not, and thus they can solve constraint satisfaction problems (CSPs). However, they cannot distinguish between the costs of the solutions, and thus cannot optimize them. For this reason, with such oracles, Grover's algorithm is required to be executed multiple times with a repeatedly updated threshold. 
        
        On the other hand, in an oracle constructed using ARC, such an oracle not only finds solutions (through the Boolean-based realizations of constraints) but also optimizes the cost (through the ARC circuit). The main properties of the ARC circuit are summarized as follows.
        \begin{itemize}
            \item If the output qubit is initialized to $\ket0$, the measurement probability of $1$ in the output of ARC \textit{increases} as the objective $C(x)$ increases for $K\geq\max \left(C(x)\right)$ (see Theorem~\ref{thm:maximization}). This proves that the oracle \textit{maximizes} the measurement probabilities of solutions based on their objective values, yielding higher probabilities for those with \textit{higher} objective values.
            \item If the output qubit is initialized to $\ket1$, the measurement probability of $1$ in the output of ARC \textit{decreases} as the objective $C(x)$ increases for $K\geq \max \left(C(x)\right)$ (see Theorem~\ref{thm:minimization}). This proves that the oracle \textit{minimizes} the measurement probabilities of solutions based on their objective values, yielding higher probabilities for those with \textit{lower} objective values.
        \end{itemize}

        Using the ARC circuit in the oracles of quantum search algorithms is a general approach to solve optimization problems.


\section{Application to State Assignment}
\label{sec:encoding}
    
    This section applies the methodology of oracle construction detailed in the previous section to the state assignment problem.

    \subsection{Oracle with ARC}
    \label{subsec:encoding-oracle}
    
\begin{figure}[!htb]
            \centering
            \includegraphics[width=\linewidth]{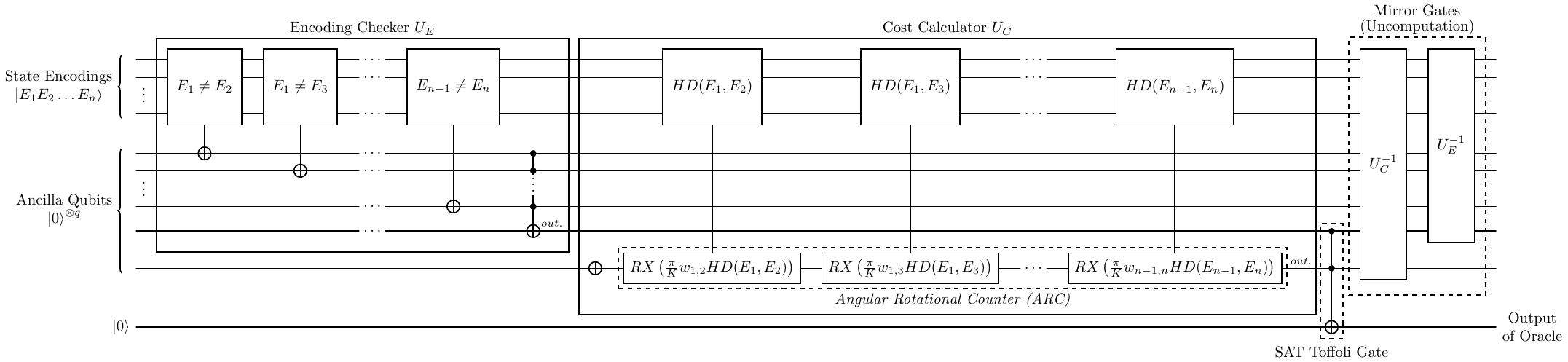}
            \caption{The quantum oracle for the state assignment problem with a cost calculator $U_C$ based on our Angular Rotational Counter (ARC), where $n$ is the number of states.}
            \label{fig:encoding-oracle}
        \end{figure}

        This subsection discusses in detail the construction of the oracle with ARC for state assignment. A full outline of the oracle is illustrated in Fig.~\ref{fig:encoding-oracle}, and each section is listed below. 
        \begin{itemize}
            \item \textbf{State Encodings ($E_1,E_2,\dots, E_n$):} These collectively form the input to the oracle. The Hadamard gates that precede the oracle in Grover's algorithm generate a superposition of all possible state encodings for a given automaton. $E_i$ represents the encoding for state $S_i$.
            
            \item \textbf{Encoding Checker ($U_E$):} This circuit outputs $\ket{1}$ if all state encodings $E_1,E_2,\dots, E_n$ are distinct. This is achieved by checking every pair of state encodings. Note that it is crucial to check that the encodings are distinct for each state because Grover's algorithm generates a superposition of all possible state encodings with Hadamard gates.
            
            \item \textbf{Cost Calculator ($U_C$):} This circuit rotationally calculates the total cost, based on a weighted sum of the Hamming distance (HD) between each pair of state encodings. This circuit is an Angular Rotational Counter (ARC).
            \item \textbf{SAT Toffoli Gate:} This gate is used to verify that all constraints are satisfied. The constraints are to have distinct encodings (output of $U_E$) and to have optimized cost (output of $U_C$). The target qubit of this Toffoli gate is the final output of the oracle.
            \item \textbf{Mirror Gates ($U_C^{-1}$ and $U_E^{-1}$):} These gates are used to reverse (uncompute) the input and ancilla qubits back to their initial state for the diffusion operator and measurement.
        \end{itemize}
    
        \subsubsection{ARC as the Cost Calculator}
        \label{subsubsec:encoding-oracle-calculator}

\begin{figure}[!htb]
            \centering
            \includegraphics[width=0.9\linewidth]{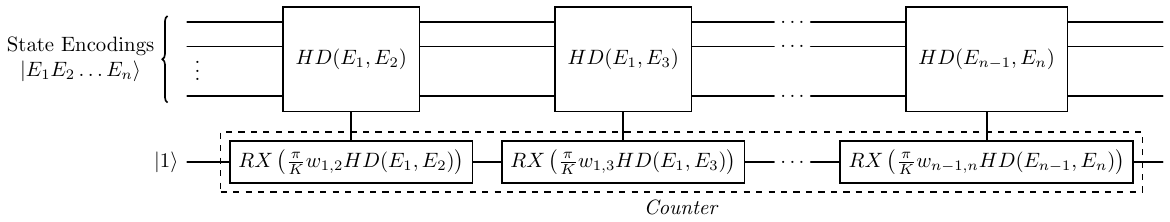}
            \caption{The ARC-based cost calculator in the oracle.}
            \label{fig:cost-calculator}
        \end{figure}

         A major part of this quantum oracle is the circuit for calculating the cost of the encoding. The construction of the cost calculator as an ARC circuit is detailed in this section. Fig.~\ref{fig:cost-calculator} illustrates the use of ARC for counting the cost $C$, based on Eq.~\eqref{eq:whd-cost}, where $w_{i,j}$ represents the weight for $HD(E_i, E_j)$, which is the HD between $E_i$ and $E_j$. The net rotation from the controlled-$RX$ gates is $$\sum_{i,j\in\{1,2,\dots,n\},i<j}\left(\frac{\pi}{K}w_{i,j} HD(E_i,E_j)\right)=\frac{\pi}{K}C.$$

        The cost calculator circuit is composed of multiple circuits that calculate the weighted HD between state encodings. Each of these circuits corresponds to two state encodings: $E_i$ and $E_j$. These circuits calculate the HD between these encodings and weigh it by $w_{i,j}$. The HD between these two encodings was defined in Eq.~\eqref{eq:hd}. The following steps are taken to calculate the HD using an ARC circuit. 
        \begin{itemize}
            \item For each index $k \in \{1,2,\dots,m\}$, $E_{i,k}$ will be compared to $E_{j,k}$.
            \item If the values are unequal, then a rotation of $\frac{\pi}{K}w_{i,j}$ (corresponding to the weight $w_{i,j}$) is applied to the counter.
            \item The final angle of the counter is $\frac{\pi}{K} w_{i,j}HD(E_i,E_j)$, which corresponds to the term $w_{i,j}HD(E_i,E_j)$.
        \end{itemize}

        \begin{figure}[!htb]
            \centering
            \includegraphics[width=0.35\linewidth]{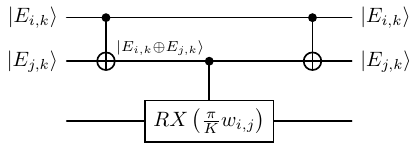}
            \caption{A component of the circuit for calculating HD. It checks for inequality between $E_{i,k}$ and $E_{j,k}$.}
            \label{fig:hd-circuit-part}
        \end{figure}

        \begin{figure}[!htb]
            \centering
            \includegraphics[width=0.4\linewidth]{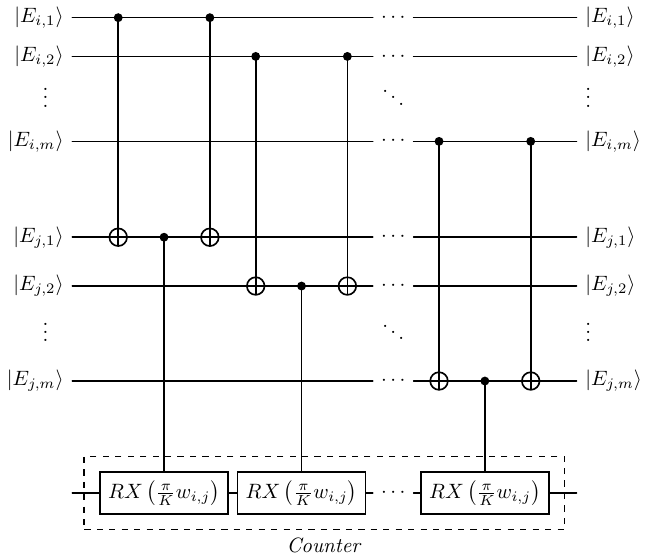}
            \caption{The circuit for calculating the weighted HD with weight $w_{i,j}$ between the state encodings $E_i$ and $E_j$.}
            \label{fig:hd-circuit}
        \end{figure}

        When comparing bits $E_{i,k}$ and $E_{j,k}$, the expression $E_{i,k} \oplus E_{j,k}$ represents $E_{i,k} \neq E_{j,k}$. This expression can be realized in a quantum circuit through CNOT gates. The qubit with state $\ket{E_{i,k}\oplus E_{j,k}}$ serves as a control for the rotational gate, which will add a rotation of $\frac{\pi}{K}w_{i,j}$ radians to the counter if $E_{i,k} \oplus E_{j,k}=1$, as illustrated in Fig.~\ref{fig:hd-circuit-part}. A second CNOT gate is then added in order to return the qubit with state $\ket{E_{i,k}\oplus E_{j,k}}$ to its initial state of $\ket{E_{j,k}}$. This circuit can be repeated for all $k \in \{1,2,\dots,m\}$ to create the circuit that calculates the total weighted HD between two encodings, as shown in Fig.~\ref{fig:hd-circuit}. The net rotation created by the controlled-$RX$ gates is $\frac{\pi}{K}w_{i,j}\sum_{k\in\{1,2,\dots,m\}}(E_{i,k}\oplus E_{j,k})=\frac{\pi}{K}w_{i,j}HD(E_i,E_j)$. This circuit is repeated for each pair of states to construct the full cost calculator.

    \subsection{Optimization Algorithm}
    \label{subsec:encoding-algorithm}
    
        This section formalizes the algorithm for state assignment using an oracle with ARC. For this algorithm, we use our developed controlled-diffuser operator $CU_s$, as discussed in our research~\cite{al-bayaty2024diffuser}. The algorithm, in words, is explained below. The pseudocode of this algorithm is detailed in Algorithm~\ref{alg:encoding}. A similar approach based on our BHT-QAOA~\cite{al-bayaty2024bht-qaoa,al-bayaty2025bht-qaoa} can be achieved, also as a classical-quantum algorithm. Example~\ref{ex:dk15} demonstrates a simulated execution of this algorithm.
        \begin{itemize}
            \item Create the oracle $U_f$.
            \begin{itemize}
                \item Initialize the circuit for $U_f$ with the following qubits.
                \begin{itemize}
                    \item $p=n\lceil\log_2{n}\rceil$ superposed input qubits, where $n$ is the number of states.
                    \item $q$ ancilla qubits that are all initialized to $\ket{0}$.
                    \item One output qubit that is initialized to $\ket{0}$.
                \end{itemize}
                \item Create the encoding checker $U_E$.
                \item Create the cost calculator $U_C$.
                \begin{itemize}
                    \item Set $K \geq \max\left(C\right)$.
                \end{itemize}
                \item Create the SAT Toffoli gate.
                \item Mirror (uncompute) gates.
                \begin{itemize}
                    \item Add the inverse of the cost calculator $U_C^{-1}$.
                    \item Add the inverse of the encoding checker $U_E^{-1}$.
                \end{itemize}
            \end{itemize}
            \item Grover's algorithm with the created oracle $U_f$ is then executed with the query complexity $O\left(\sqrt{N}\right)$, where $N=2^p$. Then, the input qubits are measured.
            \item Choose $r\geq1$ solutions based on the highest measurement probabilities.
            \item Check solutions.
        \end{itemize}

        \begin{algorithm}[!htb]
            \caption{State Encoding Using an Oracle with ARC in Grover's Algorithm}
            \label{alg:encoding}
            \begin{algorithmic}
                \State{\textbf{Input:} finite automaton $M$, weights $W$ ($w_{i,j}$ for $i,j\in\{1,2,\dots,n\}$ where $i<j$), and $r\in\mathbb{N}$}
                \State{\textbf{Output:} state encodings $E_1,E_2,\dots,E_n$} 
                \Function{$U_f$}{$M,W,K$} \Comment{Constructs the circuit for the oracle $U_f$ with $K$}
                    \State{Initialize $U_f$}
                    \State{Add encoding checker $U_E$ to $U_f$}
                    \State{Add cost calculator $U_C$ to $U_f$}
                    \State{Add SAT Toffoli gate to $U_f$}
                    \State{Add mirrors $U_C^{-1}$ and $U_E^{-1}$ to $U_f$}
                    
                    \Return{$U_f$}
                \EndFunction
                \Function{encode}{$M,W, K,r$}
                    \State{solutions $\gets$ generate $r$ from Grover's algorithm with $U_f(M,W,K)$}
                    \ForAll{solution $\in$ solutions}
                        \State{Check solution}
                        \If{solution is not a valid encoding}
                            \State{Remove solution from solutions}
                        \EndIf
                    \EndFor
                    \State{\textbf{return} solutions}
                \EndFunction
                \State{Initialize $K$} 
                \State{solutions $\gets$ {\footnotesize ENCODE}($M,W,K,r$)}
            \end{algorithmic}
        \end{algorithm}

        \begin{example}
    \label{ex:dk15}
        The goal is to assign state encodings to the automaton dk15, a benchmark from the LGSynth'91 logical benchmark set~\cite{yang1991lgsynth}. The weights calculated based on the number of transitions are listed in Table~\ref{tab:dk15-weights}. These weights are used to construct the ARC-based cost calculator illustrated in Fig.~\ref{fig:dk15-cost-calculator} (with $K=50$).

        \begin{table}[!htb]
            \centering
            \begin{tabular}{|c||c|c|c|c|c|c|}
                \hline
                $i,j$ & 1,2 & 1,3 & 1,4 & 2,3 & 2,4 & 3,4 \\
                \hline
                $w_{i,j}$ & 3 & 7 & 3 & 6 & 3 & 3 \\
                \hline
            \end{tabular}
            \caption{The transition-based weights for each pair of states $S_i$ and $S_j$ in the automaton dk15.}
            \label{tab:dk15-weights}
        \end{table}

        \begin{figure}[!htb]
            \centering
            \includegraphics[width=\linewidth]{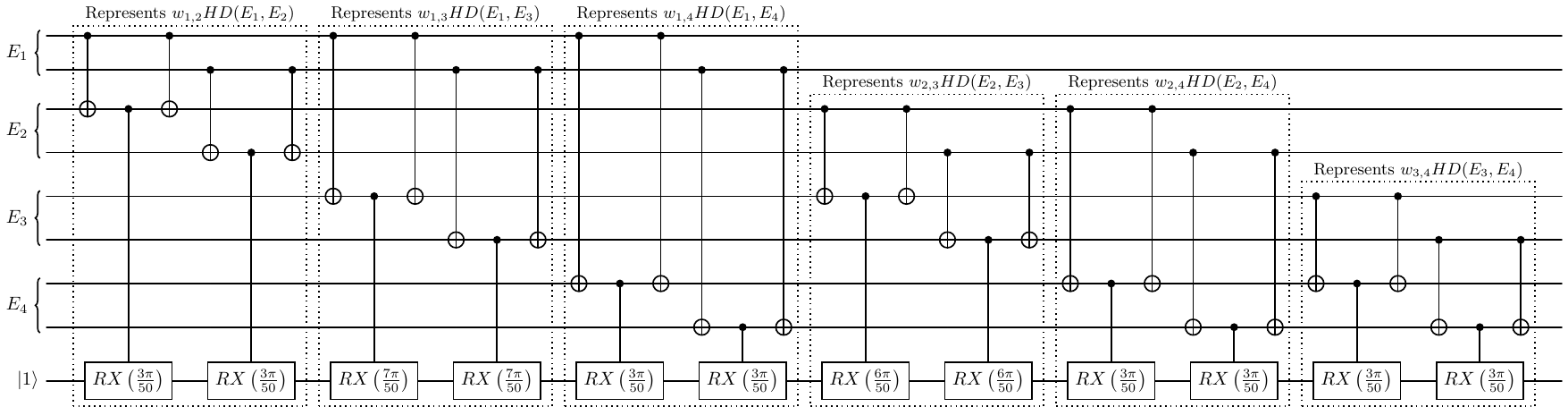}
            \caption{The cost calculator in the oracle for encoding the states of dk15.}
            \label{fig:dk15-cost-calculator}
        \end{figure}

        Grover's algorithm is executed once with an oracle constructed from the methodology introduced in this paper. The simulated result of measurement (with 2048 shots) is shown in Fig.~\ref{fig:dk15-histogram} (which was done using Qiskit~\cite{javadi2024qiskit}). The inputs that correspond to a high number of counts are solutions with minimal cost (with the cost function from Eq.~\eqref{eq:whd-cost}). The inputs that correspond to a moderately low number of counts are solutions that have a greater cost than the minimum. The inputs that correspond to an extremely low or zero count are not valid solutions. Fig.~\ref{fig:dk15-scatter} illustrates the trend of the count decreasing as the cost increases for the solutions. A fully Boolean-based based oracle would not only have a greater cost, but it would also have taken multiple executions of Grover's algorithm instead.

        \begin{figure}[!htb]
            \centering
            \includegraphics[width=\linewidth]{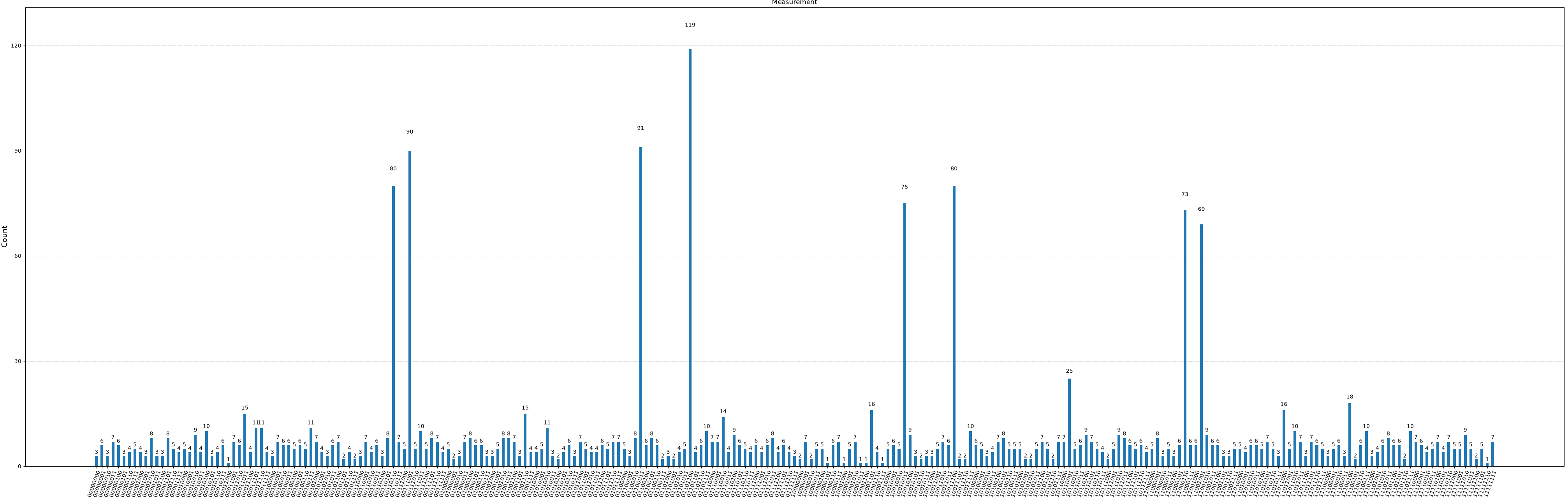}
            \caption{The cost calculator in the oracle for encoding the states of dk15.}
            \label{fig:dk15-histogram}
        \end{figure}

        \begin{figure}[!htb]
            \centering
            \includegraphics[width=.5\linewidth]{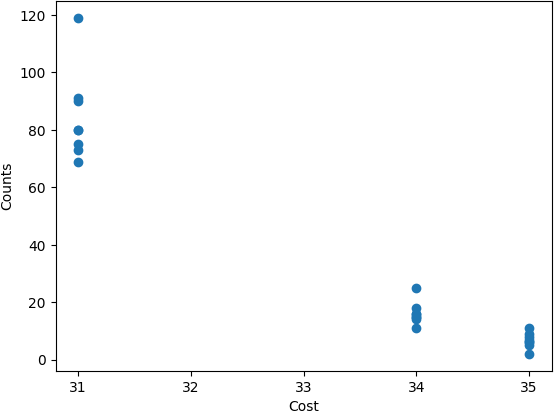}
            \caption{The corresponding number counts (shots) and cost of solutions (valid state encodings) of dk15.}
            \label{fig:dk15-scatter}
        \end{figure}

    \end{example}


\section{Experimental Methodology}
\label{sec:experiment}
    This section details how oracles with ARC for state assignment are evaluated in comparison to their corresponding Boolean-based oracles when applied to finite state machine benchmarks.

    \subsection{Benchmarks}
    \label{subsec:experiment-benchmarks}
        The 26 finite state machine benchmarks used are lion, train4, dk15, mc, tav, s8, bbtas, s27, dk27, beecount, dk14, shiftreg, dk17, ex6, ex5, lion9, ex3, ex7, bbara, train11, modulo12, ex4, dk512, ex2, donfile, and dk16, all of which are from the LGSynth'91 library~\cite{yang1991lgsynth} in the .kiss2 file format. Information on the number of states, number of inputs, and minimum encoding length of each of these benchmarks is listed in Table~\ref{tab:benchmarks}. For both the oracles with ARC and Boolean-based oracles, the weights are based on the total number of transitions between each pair of states. 

        \begin{table}[!htb]
            \centering
            \resizebox{\linewidth}{!}{
            \begin{tabular}{|l||c|c|c|c|c|c|c|c|c|c|c|c|c|}
                \hline
                & lion & train4 & dk15 & mc & tav & s8 & bbtas & s27 & dk27 & beecount & dk14 & shiftreg & dk17  \\
                \hline
                No. of States & 4 & 4 & 4 & 4 & 4 & 5 & 6 & 6 & 7 & 7 & 7 & 8 & 8 \\
                No. of Inputs & 2 & 2 & 3 & 3 & 4 & 4 & 2 & 4 & 1 & 3 & 3 & 1 & 2 \\
                Min. Enc. Len. & 2 & 2 & 2 & 2 & 2 & 3 & 3 & 3 & 3 & 3 & 3 & 3 & 3 \\
                \hline
                \hline
                & ex6 & ex5 & lion9 & ex3 & ex7 & bbara & train11 & modulo12 & ex4 & dk512 & ex2 & donfile & dk16 \\
                \hline
                No. of States & 8 & 9 & 9 & 10 & 10 & 10 & 11 & 12 & 14 & 15 & 19 & 24 & 27 \\
                No. of Inputs & 5 & 2 & 2 & 2 & 2 & 4 & 2 & 1 & 6 & 1 & 2 & 2 & 2 \\
                Min. Enc. Len. & 3 & 4 & 4 & 4 & 4 & 4 & 4 & 4 & 4 & 4 & 5 & 5 & 5 \\
                \hline
            \end{tabular}}
            \caption{The 26 benchmarks from the LGSynth'91 library~\cite{yang1991lgsynth} used for evaluation.}
            \label{tab:benchmarks}
        \end{table}

    \subsection{Boolean-Based Oracle as Baseline}
    \label{subsec:experiment-bool}
        
\begin{figure}[!htb]
            \centering
            \includegraphics[width=\linewidth]{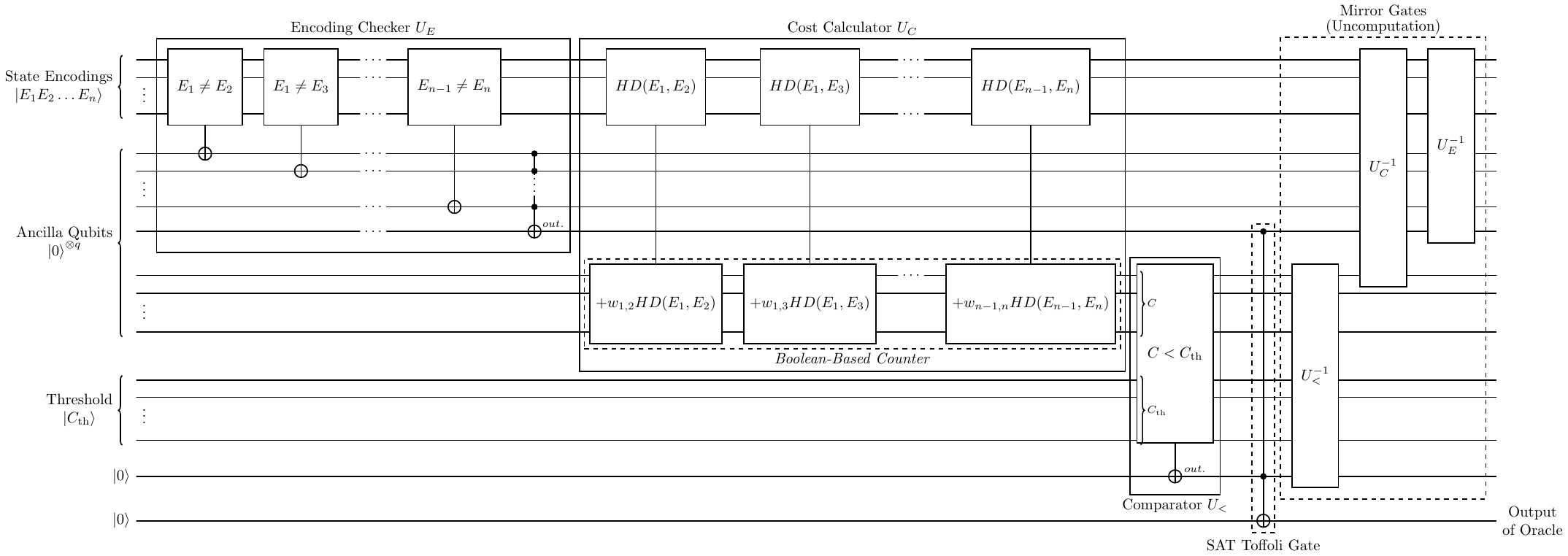}
            \caption{For comparison and evaluation, the quantum Boolean-based oracle for state assignment.}
            \label{fig:encoding-oracle-bool}
        \end{figure}

\begin{figure}[!htb]
            \centering
            \includegraphics[width=.6\linewidth]{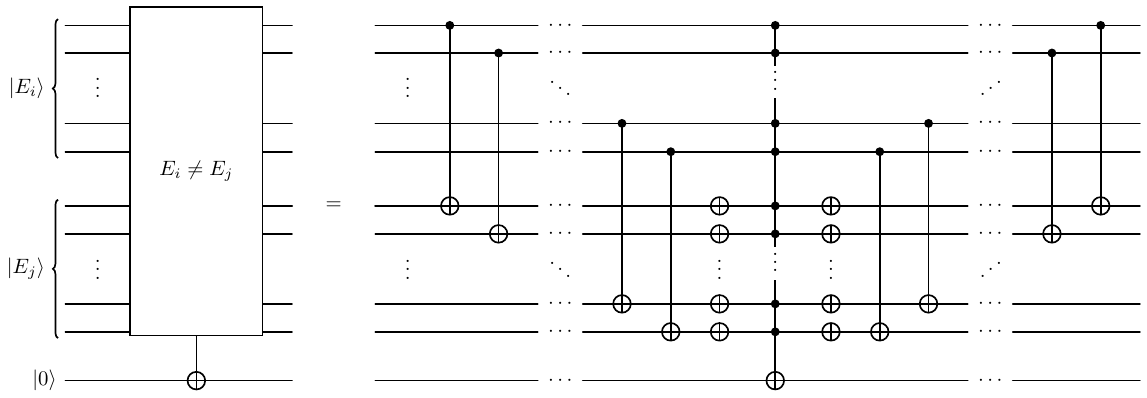}
            \captionof{figure}{The decomposition of the inequality ($\neq$) comparator used in the encoding checker $U_E$.}
            \label{fig:inequality-comparator}
        \end{figure}

        The structure of the Boolean-based oracle that is used for comparison is illustrated in Fig.~\ref{fig:encoding-oracle-bool}. Both the oracle with ARC and the Boolean-based oracle have identical Boolean-based encoding checkers (which are decomposed into the inequality comparators shown in Fig.~\ref{fig:inequality-comparator} and a multi-qubit Toffoli gate), but their cost calculators are different. As discussed earlier, the oracle with ARC uses the ARC-based cost calculator presented in Section~\ref{subsec:encoding-oracle}. 

\begin{figure}[!htb]
            \centering
            \includegraphics[width=.4\linewidth]{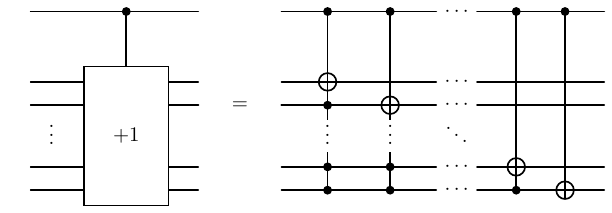}
            \captionof{figure}{The decomposition of the controlled-incrementer.}
            \label{fig:controlled-incrementer}
        \end{figure}

\begin{figure}[!htb]
            \centering
            \includegraphics[width=.6\linewidth]{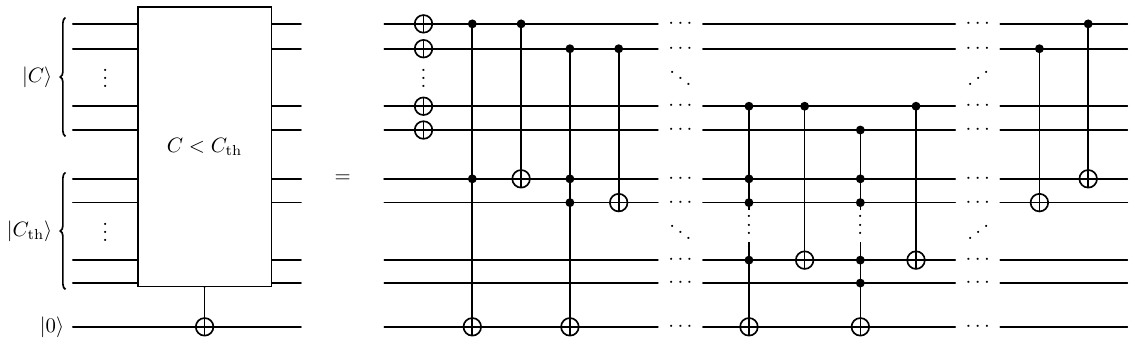}
            \captionof{figure}{The decomposition of the less-than ($<$) comparator.}
            \label{fig:less-than-comparator}
        \end{figure}

        The Boolean-based oracle instead uses a cost calculator constructed from arithmetic operators, specifically controlled-incrementer gates, that store the cost (encoded in binary) onto a multi-qubit register. In addition, this oracle also uses a comparator (which compares the cost $C$ to the threshold $C_\text{th}$). The threshold $C_\text{th}$ is stored on another multi-qubit register and is an input to the circuit. In order to optimize the cost, the threshold is updated after each execution of Grover's algorithm to the cost of the selected solution. The controlled-incrementer and less-than ($<$) comparator gates are realized from quantum reversible gates, as shown in Fig.~\ref{fig:controlled-incrementer} and Fig.~\ref{fig:less-than-comparator}, respectively.

    \subsection{Transpilation}
    \label{subsec:experiment-transpilation}
        To ensure a fair and consistent comparison of the circuit costs of oracles implemented using ARC and Boolean-based approaches, all oracles were transpiled using exactly the same configuration and settings. Specifically, the Qiskit~\cite{javadi2024qiskit} transpiler was configured with the highest optimization level (level 3), using the Qiskit Aer simulator as the target backend and the same set of basis gates: $CZ$, $RX$, $RZ$, $\sqrt{X}$, and $X$. All experiments were conducted using identical software versions: Qiskit 2.5.2 and Qiskit Aer 0.17.2. Furthermore, the transpilation process was repeated 200 times for each oracle under these same settings, and the circuit with the lowest TQC among the 200 transpiled circuits was selected for comparison.

    \subsection{Metrics and Percentage Calculation}
    \label{subsec:experiment-metrics}
        The transpiled oracles were evaluated using four metrics: qubit count, gate count, depth, and Transpilation Quantum Cost (TQC). For a specific metric, the percentage improvement of the oracle with ARC (with a value in that metric of $M_\text{ARC}$) in comparison to the Boolean-based oracle (with a value in that metric of $M_\text{Boolean}$) is $\left(\frac{M_\text{ARC}-M_\text{Boolean}}{M_\text{Boolean}}\right)\cdot100\%$. The mean percentage improvements reported were calculated from the arithmetic mean of the percentage improvements across all 26 evaluated benchmarks.


\section{Results}
\label{sec:results}

    Table~\ref{tab:comparison-boolean} lists the qubit count, gate count, depth, and Transpilation Quantum Cost (TQC) of the transpiled oracle with ARC and the transpiled Boolean-based oracle for each benchmark.
    
    \begin{table}[!htb]
    \centering
    \resizebox{\linewidth}{!}{
    \begin{tabular}{|l|rr||rrrr||rrrr||rrrr|}
        \cline{4-15}
        \multicolumn{3}{c}{} & \multicolumn{4}{|c||}{Boolean-Based Oracle} & \multicolumn{4}{c||}{Oracle with ARC} & \multicolumn{4}{c|}{\% Improvement} \\
        \hline
        Benchmark & Inputs & States & Qubits & Gates & Depth & TQC & Qubits & Gates & Depth & TQC & Qubits & Gates & Depth & TQC \\
        \hline
        lion & 2 & 4 
        & 25 & 6302 & 4453 & 10755
        & 17 & 1157 & 632 & 1789
        & 32.0\% & 81.6\% & 85.8\% & 83.4\%	\\
        train4 & 2 & 4 
        & 25 & 7087 & 5032 & 12119
        & 17 & 1229 & 662 & 1891
        & 32.0\% & 82.7\% & 86.8\% & 84.4\% \\
        dk15 & 3 & 4 
        & 29 & 51516 & 39633 & 91149
        & 17 & 1377 & 735 & 2112
        & 41.4\% & 97.3\% & 98.1\% & 97.7\% \\
        mc & 3 & 4 
        & 27 & 22806 & 17193 & 39999
        & 17 & 1230 & 664 & 1894
        & 37.0\% & 94.6\% & 96.1\% & 95.3\% \\
        tav & 4 & 4 
        & 31 & 179050 & 137010 & 316060
        & 17 & 1232 & 664 & 1896
        & 45.2\% & 99.3\% & 99.5\% & 99.4\% \\
        s8 & 4 & 5 
        & 40 & 36518 & 27233 & 63751
        & 28 & 3228 & 1673 & 4901
        & 30.0\% & 91.2\% & 93.9\% & 92.3\% \\
        bbtas & 2 & 6 
        & 48 & 46521 & 34627 & 81148
        & 36 & 4370 & 2189 & 6559
        & 25.0\% & 90.6\% & 93.7\% & 91.9\% \\
        s27 & 4 & 6 
        & 52 & 371650 & 277009 & 648659
        & 36 & 5033 & 2498 & 7531 
        & 30.8\% & 98.6\% & 99.1\% & 98.8\% \\
        dk27 & 1 & 7 
        & 57 & 48172 & 35076 & 83248
        & 45 & 6403 & 3028 & 9431
        & 21.1\% & 86.7\% & 91.4\% & 88.7\% \\
        beecount & 3 & 7 
        & 59 & 172551 & 129877 & 302428
        & 45 & 6292 & 2974 & 9266
        & 23.7\% & 96.4\% & 97.7\% & 96.9\% \\
        dk14 & 3 & 7 
        & 61 & 279337 & 206650 & 485987
        & 45 & 7173 & 3392 & 10565
        & 26.2\% & 97.4\% & 98.4\% & 97.8\% \\
        shiftreg & 1 & 8 
        & 67 & 49989 & 35862 & 85851
        & 55 & 8391 & 3818 & 12209
        & 17.9\% &	83.2\%	& 89.4\% & 85.8\%	\\
        dk17 & 2 & 8 
        & 69 & 129245 & 95481 & 224726
        & 55 & 9028 & 4124 & 13152
        & 20.3\% & 93.0\% & 95.7\% & 94.1\% \\
        ex6 & 5 & 8 
        & 75 & 1868513 & 1356341 & 3224854
        & 55 & 8931 & 4074 & 13005
        & 26.7\% & 99.5\% & 99.7\% & 99.6\% \\
        ex5 & 2 & 9
        & 89 & 186014 & 138291 & 324305
        & 75 & 15748 & 9262 & 25010
        & 15.7\% & 91.5\% & 93.3\% & 92.3\% \\
        lion9 & 2 & 9
        & 87 & 76972 & 56928 & 133900
        & 75 & 13676 & 8280 & 21956
        & 13.8\% & 82.2\% & 85.5\% & 83.6\% \\
        ex3 & 2 & 10
        & 104 & 284841 & 208799 & 493640
        & 88 & 21030 & 10232 & 31262
        & 15.4\% & 92.6\% & 95.1\% & 93.7\%	\\
        ex7 & 2 & 10
        & 104 & 269771 & 198190 & 467961
        & 88 & 20123 & 9840 & 29963
        & 15.4\% & 92.5\% & 95.0\% & 93.6\% \\
        bbara & 4 & 10
        & 104 & 284615 & 209137 & 493752
        & 88 & 20406 & 10053 & 30459
        & 15.4\% &	92.8\% & 95.2\% & 93.8\% \\
        train11 & 2 & 11
        & 114 & 75950 & 55175 & 131125
        & 102 & 22970 & 11601 & 34571
        & 10.5\% & 69.8\% & 79.0\% & 73.6\% \\
        modulo12 & 1 & 12
        & 129 & 74141 & 49399 & 123540
        & 117 & 26852 & 13443 & 40295
        & 9.3\% & 63.8\% & 72.8\% & 67.4\% \\
        ex4 & 6 & 14
        & 172 & 6097113 & 4392823 & 10489936
        & 150 & 36927 & 18268 & 55195
        & 12.8\% & 99.4\% & 99.6\% & 99.5\% \\
        dk512 & 1 & 15
        & 182 & 208106 & 145816 & 353922
        & 168 & 43813 & 21310 & 65123
        & 7.7\% & 78.9\% & 85.4\% & 81.6\% \\
        ex2 & 2 & 19
        & 287 & 941555 & 668445 & 1610000
        & 269 & 88550 & 43687 & 132237
        & 6.3\% & 90.6\% & 93.5\% & 91.8\% \\
        donfile & 2 & 24
        & 417 & 990210 & 692089 & 1682299
        & 399 & 138169 & 68714 & 206883
        & 4.3\% & 86.0\% & 90.1\% & 87.7\% \\
        dk16 & 2 & 27
        & 509 & 1733691 & 1212996 & 2946687
        & 489 & 179056 & 88206 & 267262 
        & 3.9\% & 89.7\% & 92.7\% & 90.9\% \\
        \hline
    \end{tabular}}
    \caption{Comparison of the transpiled oracle with ARC and the transpiled Boolean-based oracle for each benchmark.}
    \label{tab:comparison-boolean}
    \end{table}

    In this table, the percentage improvement in qubit count ranged from 3.9\% to 45.2\%, with a mean improvement of 20.8\%. More specifically, the size of the encoding checker grows quadratically with the number of states, whereas the cost calculator has constant size in the ARC-based approach and grows logarithmically with respect to the maximum cost in the Boolean-based approach. Nevertheless, the use of ARC consistently reduces the qubit count across all benchmarks. This reduction stems from the fact that an ARC-based oracle requires only a single-qubit register to encode the cost, whereas a Boolean-based oracle requires two multi-qubit registers: one to encode the cost and another to encode the threshold. The sizes of both registers scale with the range of objective function values, resulting in a higher overall qubit requirement for the Boolean-based approach.

    The percentage improvements range from 63.8\% to 99.5\% in gate count, 72.8\% to 99.7\% in circuit depth, and 67.4\% to 99.6\% in TQC, with corresponding mean improvements of 89.3\%, 92.4\%, and 90.6\%, respectively. These substantial reductions can primarily be attributed to the use of controlled-$RX$ gates in ARC-based oracles, as opposed to controlled-incrementer gates in their Boolean-based counterparts. When decomposed, controlled-incrementer gates require more elementary gates and result in greater circuit depth. Furthermore, ARC-based oracles eliminate the need for a comparator, which is a relatively costly component in Boolean-based oracles. By avoiding both the overhead associated with controlled-incrementer gates and the comparator, ARC significantly reduces oracle circuit costs. Overall, across all 26 evaluated benchmarks, ARC-based oracles consistently achieve lower circuit costs and require fewer qubits than their corresponding Boolean-based oracles after transpilation.
    

\section{Discussion}
\label{sec:discussion}

    This section discusses the theoretical and experimental findings of this research.

    The Angular Rotational Counter (ARC) circuit computes the objective function value $C(x)$ of a given candidate solution $x$ through rotationally encoding it with the angle $\frac{\pi}{K}C(x)$ radians onto a single-qubit register. This is an alternative, cost-effective approach to explicitly encoding $C(x)$ as a binary value onto a multi-qubit register, which was used previously in Boolean-based oracles for optimization problems. 
    
    \subsection{Circuit Cost in Comparison to Boolean-Based Methods}
    \label{subsec:discussion-cost}
        Oracles with ARC have a lower circuit cost than corresponding Boolean-based oracles for three main reasons: oracles with ARC use controlled quantum rotational gates instead of controlled-incrementer gates, do not use a comparator, and have a single-qubit register instead of multi-qubit registers. Firstly, the controlled-incrementer gates and other arithmetic operators used in a Boolean-based circuit for computing cost must be decomposed into many individual gates, such as $X$ (NOT), $CX$ (CNOT), and Toffoli gates. Such arithmetic operators increase in cost depending on the size of the multi-qubit register and are more expensive than controlled quantum rotational gates. Secondly, the Boolean-based oracle always requires a comparator, which has similar disadvantages to arithmetic operators, in order for the oracle to be able to distinguish the cost of solutions. In contrast, oracles with ARC do not use these expensive comparators because the output of ARC is already dependent on the objective function values. Thirdly, ARC uses a single-qubit register, while Boolean-based cost calculators use a multi-qubit register that increases in qubit count based on the objective function values.

    \subsection{Objective Functions and Measurement Probabilities}
    \label{subsec:discussion-objective}
        The ARC circuit is naturally suited for objective functions that can be decomposed into a weighted sum of Boolean functions, where the outputs of the Boolean functions are treated as real numbers, although they must be 0 or 1, because of the inherently logical property of controlled gates (which represent the Boolean functions) and the continuity of rotational angles (which represent the weights). Here, the applicability of ARC for weighted Hamming distance cost functions has been demonstrated by decomposing the Hamming distance function into a sum of Boolean functions. Many other functions can be decomposed in similar ways. For functions that cannot be decomposed into such a form, the ARC circuit should be modified. 
    
        The measurement probabilities of the output qubit of ARC are dependent on the objective function. However, the ARC circuit itself does not automatically yield valid or optimal solutions. A marking circuit that identifies valid solutions and filters out the rest must be added to the oracle. An oracle cannot work without such a circuit. In the context of optimization, the ARC circuit should be used within a quantum algorithm. Furthermore, for an ARC circuit to function as intended, the constant $K$ (in $\frac{\pi}{K}C(x)$ as the rotational angle) should be set to a value greater than or equal to the maximum objective function value because of the periodic nature of the sinusoidal functions that appear in rotational gates. However, it is important to note that in physical hardware, there is a limit to how small an angle can be~\cite{greenbaum2018error, lazuar2023calibration}, corresponding to a limit on how large $K$ can be. It is also important to note that for higher values of $K$, and thus smaller and more precise angles, the impact of noise will become a larger problem for performance. 

    \subsection{Complexity Analysis}
    \label{subsec:discussion-complexity}
        The exact time complexity of an ARC circuit depends on the objective function. A general formula for the complexity, however, can be derived, as achieved in Theorem~\ref{thm:complexity}.

        \begin{theorem}
        \label{thm:complexity}
            The time complexity $T(q)$ of an ARC circuit is determined by Eq.~\eqref{eq:complexity}, where the ARC circuit contains $q\in\mathbb{N}$ qubits, the objective function is of the form in Eq.~\eqref{eq:objective}, and $T_i(q)$ is the time complexity of the quantum circuit realization of $f_i(x)$.
            \begin{equation}
            \label{eq:complexity}
                T(q) = O\left( m + \sum_{i=1}^mT_i(q) \right)
            \end{equation}
        \end{theorem}
        \begin{proof}
            The ARC circuit contains a circuit realization of $f_i(x)$, a controlled-$RX\left(\frac{\pi}{K}w_i\right)$, and a circuit realization of $f_i^{-1}(x)$ for all $i\in\{1,2,\dots,m\}$. 
            
            The time complexity of the circuit realization of $f_i(x)$ is given as $T_i(q)$. The time complexity of its inverse, $f_i^{-1}(x)$, is also $T_i(q)$, because the circuit for $f_i(x)$ can be reversed to obtain the circuit for $f_i^{-1}(x)$. The total time complexity  of all $f_i(x)$ and $f_i^{-1}(x)$ circuits is therefore $\sum_{i=1}^m2T_i(q)$. The controlled-$RX\left(\frac{\pi}{K}w_i\right)$ gate has a constant time complexity. Let this constant be $t_{CRX}$.

            The time complexity of the whole ARC circuit $T(q)$ would then be 
            $$T(q) = O\left( \sum_{i=1}^m\left(2T_i(q) + t_{CRX}\right) \right).$$ 
            This simplifies to $T(q) = O\left( m + \sum_{i=1}^mT_i(q) \right)$
        \end{proof}

        \subsubsection{Comparison of State Assignment with ARC to Classical Algorithm}
        \label{subsubsec:discussion-complexity-encoding-classical}
            In the classical-quantum algorithm for state assignment introduced in this paper, Grover's algorithm is executed once with $n\lceil\log_2 n\rceil$ input qubits. Therefore, the query complexity of this classical-quantum algorithm is $O\left(\sqrt{n^{n}}\right)$. In contrast, an equivalent classical algorithm makes at most $\frac{2^{\lceil\log_2 n\rceil}!}{(2^{\lceil\log_2 n\rceil}-n)!}$ queries, which is the total number of possible assignments of state encodings. Accordingly, the query complexity of this algorithm is $O(n!)$, which grows faster than $O\left(\sqrt{n^{n}}\right)$. Hence, this classical-quantum algorithm introduced in this work has a lower query complexity than an equivalent classical algorithm. 

        \subsubsection{Comparison of State Assignment with ARC to Previous Work}
        \label{subsubsec:discussion-complexity-encoding-previous}
            Previous work on creating a quantum oracle for state assignment was discussed in \cite{tsai2020encoding}. Notable differences include the applicability, cost function, circuit cost, and corresponding algorithm. The cost calculator of the oracle in the previous method~\cite{tsai2020encoding} is constructed with a Boolean-based counter and a comparator (for a threshold). According to~\cite{tsai2020encoding}, the time complexity of the cost calculator circuit and comparator is $O\left(2^{3n_S}n_S^2+2^{2n_I}(2^{n_S}+n_I)(n_Sn_I)\right)$, where $n_S$ is the number of states and $n_I$ is the number of inputs. However, it is important to note that the scaling in time complexity of the many multi-qubit Toffoli gates used in the circuit is not considered in the calculation from~\cite{tsai2020encoding} and that the previous method encodes both states and inputs with a more accurate cost function.
    
            In contrast, the cost calculator of the oracle in our method is constructed with an ARC circuit. The cost calculator circuit is composed of at most $\frac{n(n-1)}{2}$ weighted HD calculator circuits, each of which contains $2\lceil\log_2n\rceil$ CNOT gates and $\lceil\log_2n\rceil$ controlled-$RX$ gates. Therefore, by applying Theorem~\ref{thm:complexity} and letting the time complexity of a CNOT and controlled-$RX$ gate be the constants $t_{CNOT}$ and $t_{CRX}$, respectively, the cost calculator circuit has a time complexity of 
            $$O\left(\frac{n(n-1)}{2}+\sum_{i=1}^\frac{n(n-1)}{2}\left(2\lceil\log_2n\rceil t_{CNOT}+\lceil\log_2n\rceil t_{CRX}\right)\right)
            =O\left(n^2\log n\right).$$ 
            While the cost calculator circuit in the previous method has a time complexity that is more than exponential, the cost calculator circuit in our method has a time complexity that is quadratic-logarithmic, which is significantly lower than the previous method.

    \subsection{Future Directions}
    \label{subsec:discussion-future}
        There are several promising directions in which this research can be extended. Firstly, future work could investigate the application of the ARC circuit to a broader class of objective functions and optimization problems beyond the finite state machine benchmarks considered in this study. In particular, exploring different objective function representations may reveal additional opportunities for constructing cost-effective, resource-efficient quantum oracles. One interesting direction is to investigate rotational encodings of objective function values as an alternative to explicit binary representations in more general contexts.
        
        Secondly, the ARC circuit could be further developed and integrated into other quantum algorithms. Investigating its applicability to different quantum optimization frameworks and oracle-based algorithms may help establish the broader utility of the proposed approach.
        
        Thirdly, further research could focus on improving the query and time complexities of optimization algorithms that employ ARC-based oracles. Specifically, theoretical analyses and experimental evaluations could help determine whether the observed reductions in quantum circuit resources can translate into improvements in overall algorithmic efficiency.
        
        Finally, the scalability of ARC-based implementations warrants further investigation, particularly for automata with larger state spaces. Evaluating larger problem instances using more powerful computational platforms would provide additional insight into their resource requirements and practical limitations. Such investigations could help establish the applicability of ARC to increasingly complex optimization problems.


\section{Conclusions}
\label{sec:conclusions}

    This paper introduces the Angular Rotational Counter (ARC), a quantum circuit design that encodes decomposable objective functions through the accumulation of quantum rotational operations on a single-qubit register. We establish the correctness of the proposed circuit, formally characterize the quantum state of its output qubit for candidate solutions, and demonstrate its application to constructing oracles for quantum search algorithms in optimization problems. Notably, we prove that the measurement probabilities of the ARC output qubit vary monotonically with the objective function values. This property enables the construction of oracles that favor candidate solutions according to either the maximization or minimization of the objective, depending on the initialization of the input qubits in the ARC circuit.
    
    To investigate the practical applicability of ARC, we employ it to construct oracles for the state assignment problem in finite automata and state machines, using a heuristic weighted Hamming distance as the objective function. We evaluate the resulting oracles with ARC against their conventional Boolean-based counterparts across 26 finite state machine benchmarks using four quantum circuit cost metrics. The experimental results demonstrate substantial reductions in circuit resource requirements and resulting costs. Specifically, the implementations with ARC achieve arithmetic mean reductions of 20.8\% in qubit count, 89.3\% in gate count, 92.4\% in circuit depth, and 90.6\% in transpiled circuit cost. These results highlight the potential of rotational encoding to substantially reduce the quantum circuit resources required for oracle construction compared with conventional Boolean-based representations.
    
    Although these results demonstrate significant improvements in quantum circuit cost and resource efficiency, they do not, by themselves, establish corresponding reductions in physical execution time or an asymptotic quantum speedup. Nevertheless, the consistent cost reductions observed across the evaluated benchmarks provide encouraging evidence for the practical utility of ARC-based oracle construction. More broadly, this work motivates further investigation into rotational encodings of function values as an alternative to explicit binary representations in quantum circuits. The proposed ARC framework also provides a foundation for exploring oracle constructions for a broader range of objective functions and optimization problems. As quantum computing continues to evolve, our results suggest that rotational encoding, exemplified by resource-efficient oracle architectures such as ARC, may offer a promising alternative design framework for quantum oracles, paving the way toward more practical and scalable quantum optimization algorithms.


\section*{Acknowledgments}
\label{sec:acknowledgments}
Computational resources used in this work were supported by the Oregon Regional Computing Accelerator (Orca), funded by NSF CC*-2346732.


\bibliographystyle{ieeetr}
\bibliography{sources}

\end{document}